\documentclass[aps,prb,reprint,groupedaddress,twoside,showpacs]{revtex4-1}
\usepackage{amsfonts}
\usepackage{amsmath}
\usepackage{amssymb}
\usepackage{graphicx}
\usepackage{CJK}
\usepackage{mathptmx}
\usepackage[breaklinks,colorlinks,linkcolor=blue,citecolor=blue,urlcolor=black,bookmarks=false]{hyperref}

\input tcilatex.tex
\begin{document}

\begin{CJK*}{Bg5}{bsmi}
\title{Efficient quantum transport simulation for bulk graphene heterojunctions}
\author{Ming-Hao Liu (¼B©ú»¨)}
\email{Corresponding author: minghao.liu.taiwan@gmail.com}
\affiliation{Institut f\"{u}r Theoretische Physik, Universit\"{a}t Regensburg, D-93040
Regensburg, Germany}
\author{Klaus Richter}
\affiliation{Institut f\"{u}r Theoretische Physik, Universit\"{a}t Regensburg, D-93040
Regensburg, Germany}
\pacs{72.80.Vp,73.23.Ad,73.40.Gk,72.10.Bg}
\begin{abstract}
The quantum transport formalism based on tight-binding models is known to be
powerful in dealing with a wide range of open physical systems subject to
external driving forces but is, at the same time, limited by the memory requirement's
increasing with the number of atomic sites in the scattering region.
Here we demonstrate how to achieve an accurate simulation of
quantum transport feasible for experimentally sized bulk graphene heterojunctions
at a strongly reduced computational cost. Without free tuning parameters, we show excellent
agreement with a recent experiment on Klein backscattering
[A.\ F.\ Young and P.\ Kim, Nature Phys.\ \textbf{5}, 222 (2009)].
\end{abstract}
\date{\today}
\maketitle
\end{CJK*}

\section{Introduction}

Electronic transport is one of the important fields among the increasing
number of fundamental studies\cite{CastroNeto2009,DasSarma2011} of graphene,
a one-atom-thick carbon honeycomb lattice.\cite{Novoselov2004} Due to the
gapless and chiral nature of its electronic structure, graphene exhibits
energy dispersions linear in momentum, the transport carriers behave like
massless Dirac fermions, and the properties based on Schr\"{o}dinger wave
mechanics in semiconductor physics have to be retreated by Dirac-type
physics in graphene. Tunneling across \emph{pn} and \emph{pnp} junctions is
perhaps the most popular example that shows how different the charge
carriers behave, compared to semiconductor heterostructures. By solving the
Dirac equation, perfect transmission at normal incidence across a potential
step\cite{Cheianov2006} as well as a potential barrier\cite{Katsnelson2006}
was shown for monolayer graphene. This mimicks the Klein paradox in quantum
electrodynamics \cite{Klein1929} and was later referred to as Klein
tunneling,\cite{Beenakker2008,Allain2011} which attracted both experimental%
\cite%
{Huard2007,Williams2007,Ozyilmaz2007,Liu2008c,Gorbachev2008,Stander2009,Young2009,Nam2011}
and further theoretical\cite%
{Zhang2008,Shytov2008,Gorbachev2008,Low2009a,Yamakage2009,Low2009,Rossi2010,Masir2010,Liu2012}
investigations.

The Dirac theory, an effective approach valid only for low-energy
excitations, generally serves as a starting point for theoretical studies of
transport in graphene and can often provide analytical results to capture
basic physical insights for certain problems with simplified system
geometries. For further considerations, such as to maintain the lattice
information on graphene or to account for complicated geometries and more
realistic factors, one has to resort to more advanced theoretical models.
The tight-binding model (TBM), a commonly used semiemperical approach for
electronic structure calculations in solid state physics,\cite{Grosso2000}
allows for consideration of more complete band information on graphene at a
low computational cost. The combination of the TBM with nonequilibrium
Green's function approaches forms the modern quantum transport formalism,%
\cite{Datta1995} which is able to deal with a wide range of conductors
composed of a scattering region and external leads with or without bias.
\begin{figure}[b]
\includegraphics[width=\columnwidth]{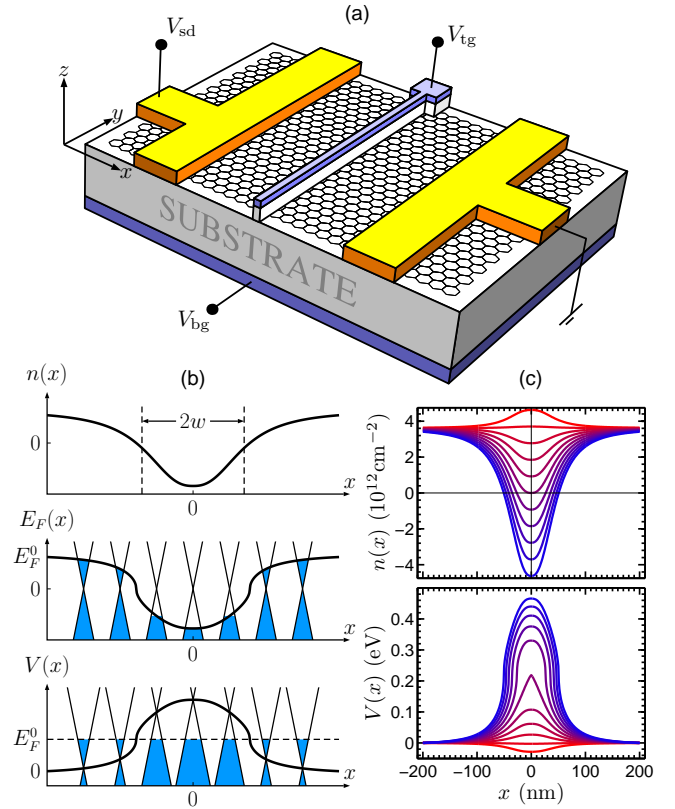}
\caption{(Color online) (a) Schematic of double-gated graphene. (b) Carrier
density profile $n(x)$ (top) and its corresponding local Fermi level $%
E_{F}(x)$ (middle). The extracted potential profile $V(x)$ (bottom) is given
by the difference between the global Fermi level $E_{F}^{0}$ and $E_{F}(x)$;
see text. (c) Reproduced densities $n(x)$ provided in the Supplementary
Material for Ref.\ \onlinecite{Young2009} with $V_{\text{bg}}=50\unit{V}$
and $V_{\text{tg}}=-8.9,-7.9,\cdots ,0.1,1.1\unit{V}$ (curves from bottom to
top), and the extracted corresponding $V(x)$ (curves from top to bottom).}
\label{fig1}
\end{figure}
The description of the graphene scattering region of interest, however,
requires a TBM Hamiltonian matrix,%
\begin{equation}
H_{\text{gnr}}\left( V,t,t^{\prime }\right) =\sum_{n=1}^{N}V_{n}c_{n}^{\dag
}c_{n}-t\sum_{\langle m,n\rangle }c_{n}^{\dag }c_{m}-t^{\prime
}\sum_{\langle \!\langle m,n\rangle \!\rangle }c_{n}^{\dag }c_{m},
\label{Hgnr}
\end{equation}%
whose matrix size depends on the involved number of atomic sites $N$ and
therefore imposes a computational limit when addressing realistic
experimental system sizes. This is partly the reason why many quantum
transport studies address graphene \textquotedblleft
nanoribbons\textquotedblright\ rather than large-area\ graphene. The
notation in Eq.\ \eqref{Hgnr} is described as follows: $t$ ($t^{\prime }$)
is the nearest (next nearest) neighbor hopping parameter, $V_{n}$ is the
local potential energy at site $n$, $c_{n}^{\dag }$ ($c_{n}$) creates
(annihilates) a charge carrier at the $n$th site, and the summation $%
\sum_{\langle m,n\rangle }$ ($\sum_{\langle \!\langle m,n\rangle \!\rangle }$%
) runs over all $m$ and $n$ site indices that are nearest (next nearest) to
each other within the scattering region.

Typical sizes of graphene flakes for experimental transport investigations
amount to a few microns by a few microns, but even a $1\unit{%
\mu%
m}\times 1\unit{%
\mu%
m}$ graphene flake contains roughly $10^{7}$ atoms, leading to a spinless
single-orbital TBM Hamiltonian matrix of more than $10^{14}$ elements that
requires an exceeding memory and hence an unreasonable computation burden.
TBM-based quantum transport for bulk materials therefore requires further
improvements to overcome the issue of the limited scattering region size. In
this paper, we demonstrate how an accurate TBM-based transport calculation
for bulk graphene heterojunctions can be performed without free parameters,
circumventing the problem of large system scales.

To achieve such a TBM bulk transport simulation, two crucial concepts are
required, namely, extraction of a realistic potential profile and
description of a bulk graphene scattering region, which are described in
Sec.\ \ref{sec theory}, where a brief summary of the quantum transport
formalism is also included (Sec.\ \ref{sec qt}). In Sec.\ \ref{sec exp}, we
revisit and simulate the recent Klein backscattering experiment\cite%
{Young2009} for transport through double-gated graphene [as depicted in
Fig.\ \ref{fig1}(a)] to compare with and to demonstrate our approach.
Section \ref{sec summary} summarizes the present work.

\section{Theoretical formulation\label{sec theory}}

\subsection{Extraction of a realistic potential profile}

A theoretical study of transport in graphene, whether based on Dirac theory
or the TBM formalism, requires the potential $V(x)$ as an input, which
actually means the local energy offset of the Dirac point and is often
regarded directly as the electric potential. In fact, the application of a
gate voltage $V_{g}$ does not directly raise the Dirac cone by $-eV_{g}$ ($%
-e $ being the electron charge) but enhances or depletes the carrier
density, hence raising or lowering the local Fermi level. For double-gated
graphene [Fig.\ \ref{fig1}(a)], the combination of a top-gate voltage $V_{%
\text{tg}}$ and a back-gate voltage $V_{\text{bg}}$ results in a carrier
density profile $n(x)$ such as that shown in the upper panel in Fig.\ \ref%
{fig1}(b). Its energy dependence, $n(E)=\limfunc{sgn}\left( E\right)
E^{2}/[\pi (\hbar v_{F})^{2}]$, is obtained by integrating the density of
states over energy. Defining the local Fermi level as%
\begin{equation}
E_{F}(x)=\limfunc{sgn}[n(x)]\hbar v_{F}\sqrt{\pi |n(x)|},  \label{EF(x)}
\end{equation}%
one obtains the spatially varying height of the filled states, as depicted
in the middle panel in Fig.\ \ref{fig1}(b). In a transport calculation, the
global Fermi level $E_{F}^{0}$ is a fixed quantity. Hence to account for the
profiles of $E_{F}(x)$ and $n(x)$, one shifts the local band offset by
applying a local potential%
\begin{equation}
V(x)=E_{F}^{0}-E_{F}(x),  \label{V(x)}
\end{equation}%
as depicted in the lower panel in Fig.\ \ref{fig1}(b). This completes the
extraction of the potential profile from the carrier density profile. Note
that the above model makes use of the linear density of states that is
normally valid in the experimental range of the carrier density, although
the energy dispersion based on the TBM covers the full range. The energy
range beyond the Dirac model with a nonlinear density of states can, in
principle, be treated within the TBM similarly to the process introduced
above, but this would be relevant only far from the energy range of interest.

A realistic carrier density profile depends on the experimental geometry and
dielectric material of the gate fabrication. In the experiment in Ref.\ %
\onlinecite{Young2009}, $n(x)$ was obtained from an electrostatic simulation
and empirically described by%
\begin{equation}
n(x)=\left( \frac{12.8V_{\text{tg}}}{1+\left\vert x/w\right\vert ^{2.5}}+V_{%
\text{bg}}\right) C_{\text{bg}},  \label{n(x)}
\end{equation}%
where $12.8$ accounts for the effectiveness of the top-gate relative to the
back-gate, $C_{\text{bg}}\approx 7.23\times 10^{10}\unit{cm}^{-2}/\unit{V}$
is the classical (electron number) capacitance of a $290\unit{nm}$-thick SiO$%
_{2}$ substrate, and the effective half width of the top-gate is $w=46\unit{%
nm}$.\cite{Young2009} Figure \ref{fig1}(c) shows various carrier density
profiles described by Eq.\ \eqref{n(x)}, subject to $V_{\text{bg}}=50\unit{V}
$ and various $V_{\text{tg}}$, and the extracted potential profiles, Eqs.\ %
\eqref{EF(x)} and \eqref{V(x)}.

\subsection{Bulk graphene scattering region\label{sec tbm}}

In band theory, the electronic structure of a crystal lattice can be solved
by applying the Bloch theorem, which allows us to reduce the problem with
infinitely repeated unit cells to only one due to translation invariance
along each space dimension. For transport calculations, however, the
scattering region of interest is composed of a certain finite-size area and
is generally not translationally invariant. For a large flake of
double-gated graphene, such as that sketched in Fig.\ \ref{fig1}(a), the
transverse dimension (along $y$) is typically a few microns in width so that
the edges are of minor importance, and we can then assume translational
invariance in the $y$ direction.
\begin{figure*}[t]
\includegraphics[width=0.7\textwidth]{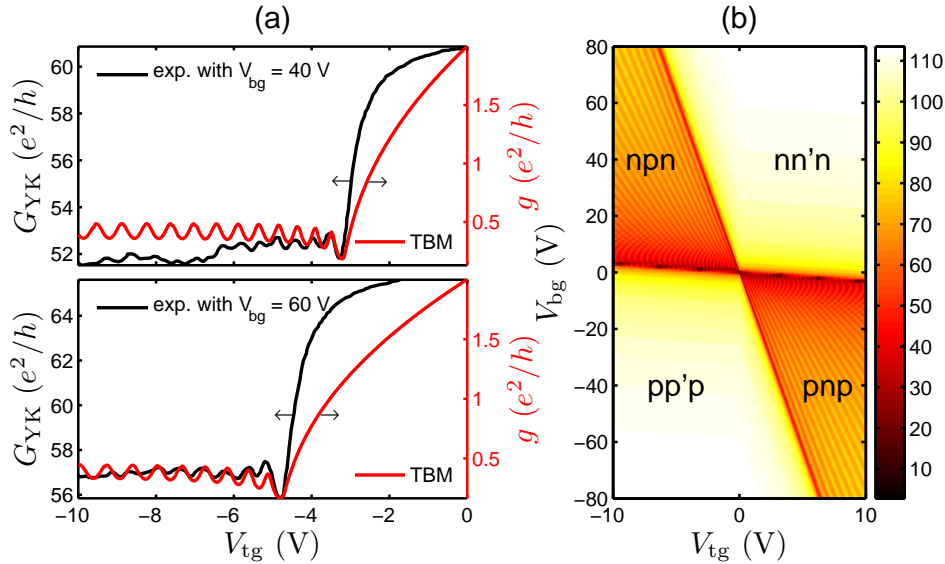}
\caption{(Color online) (a) Comparison of the top-gate voltage dependence of
the measured conductance $G_{\text{YK}}$ \protect\cite{Note1} and the
computed single-mode conductance $g$ at $V_{\text{bg}}=40\unit{V}$ and $V_{%
\text{bg}}=60\unit{V}$. (b) Conductance map of $G(V_{\text{tg}},V_{\text{bg}%
})$.}
\label{fig2}
\end{figure*}

Consider bulk graphene oriented with zigzag carbon chains along the $x$
direction. Up to nearest neighbor hopping, the minimal unit cell can be
chosen as one hexagon row, i.e., a graphene nanoribbon with zigzag chain
number $N_{z}=2$ with transverse periodicity $W=3a$, $a\approx 1.42\unit{%
\text{\AA}%
}$ being the bond length. The wave function at the bottom site $\langle
x,y_{B}|\varphi \rangle $ of the unit cell is related to that at the top
site $\langle x,y_{T}|\varphi \rangle $ through the Bloch theorem as\cite%
{Wimmer2008} $\langle x,y_{T}+a|\varphi \rangle =e^{ik_{y}W}\langle
x,y_{B}|\varphi \rangle $, implying $|x,y_{T}\rangle \langle x,y_{T}+a|$ $%
=e^{ik_{y}W}|x,y_{T}\rangle \langle x,y_{B}|$, where $k_{y}$ is the Bloch
momentum defined within $k_{y}W\in \lbrack -\pi ,\pi ]$. This means that a
kinetic hopping across the upper boundary of the unit cell $|x,y_{T}\rangle
\langle x,y_{T}+a|$ can be equivalently expressed as a periodic hopping $%
|x,y_{T}\rangle \langle x,y_{B}|$ modulated by the phase $e^{ik_{y}W}$
arising from the Bloch theorem. Similarly, one can obtain for the lower
boundary $|x,y_{B}\rangle \langle x,y_{B}-a|=e^{-ik_{y}W}|x,y_{B}\rangle
\langle x,y_{T}|$. Incorporating these periodic hopping terms, the TBM
Hamiltonian for a bulk graphene scattering region can therefore be written as%
\begin{align}
H_{\text{bulk}}(V,t;k_{y})& =H_{\text{gnr}}(V,t,0)  \notag \\
& +\left( -te^{ik_{y}W}\sum_{m}c_{T_{m}}^{\dag }c_{B_{m}}+\text{H.c.}\right)
,  \label{Hbulk}
\end{align}%
where $c_{T_{m}}^{\dag }$ ($c_{B_{m}}$) creates (annihilates) a charge
carrier at the top (bottom) edge site of the $m$th hexagon along $x$, and $%
H_{\text{gnr}}(V,t,0)$, given in Eq.\ \eqref{Hgnr}, describes an $N_{z}=2$
graphene nanoribbon. Note that the above description for a bulk scattering
region is restricted neither to nearest neighbor hopping ($t^{\prime }=0$)
nor to the material graphene. For the present bulk transport simulation,
however, next-nearest-neighbor hopping does not play an important role and
we adopt $t=3\unit{eV}$ and $t^{\prime }=0$ throughout Sec.\ \ref{sec exp}.

\subsection{Quantum transport formalism\label{sec qt}}

The quantum transport simulation in the present work is restricted to the
linear response regime at zero temperature. Thus the Landauer conductance%
\begin{equation}
g(E_{F}^{0})=\frac{e^{2}/h}{2k_{F}}%
\int_{-k_{F}}^{k_{F}}T(E_{F}^{0};k_{y})dk_{y}  \label{g}
\end{equation}%
is the main object and is obtained by integrating the transmission function%
\begin{equation}
T(E;k_{y})=\func{Tr}(\Gamma _{R}G_{R}\Gamma _{L}G_{R}^{\dag }),  \label{T}
\end{equation}%
which is equivalent to the Fisher-Lee relation.\cite{Fisher1981} The Fermi
wave vector in Eq.\ \eqref{g} is approximated from the low-energy linear
dispersion by $k_{F}=E_{F}^{0}/(\hbar v_{F})=E_{F}^{0}/(3ta/2)$. Note that
the spin degeneracy is neglected here, while the valley degeneracy is
inherently incorporated in $H_{\text{bulk}}$.

The retarded Green's function $G_{R}$ of the scattering region at energy $E$
in Eq.\ \eqref{T} is obtained from%
\begin{equation}
G_{R}(E;k_{y})=\frac{1}{E-[H_{\text{bulk}}(V,t;k_{y})+\Sigma _{L}+\Sigma
_{R}]},  \label{GR}
\end{equation}%
where $H_{\text{bulk}}(V,t;k_{y})$ has been given in Eq.\ \eqref{Hbulk} and $%
\Sigma _{L}$ ($\Sigma _{R}$) is the self-energy due to the left (right) lead
composed of a semi-infinite repetition of unit cells. Adopting a
Schur-decomposition-based algorithm for the singular hopping matrix type,%
\cite{Wimmer2008} the periodic hoppings as used in $H_{\text{bulk}}
$ can also be included in $\Sigma _{L}$ and $\Sigma _{R}$, enabling us to
study pure bulk-to-bulk transmission. The spectral matrix functions $\Gamma
_{l}$, with $l=L,R$, in Eq.\ \eqref{T} are given by $\Gamma _{l}=i(\Sigma
_{l}-\Sigma _{l}^{\dag })$.{}

\section{Klein backscattering experiment revisited\label{sec exp}}

\subsection{Gate-voltage dependence}

Now we revisit the experiment in Ref.\ \onlinecite{Young2009} by considering
the extracted realistic potential $V(x)$ and applying the bulk TBM\
transport formalism introduced above. As shown in Fig.\ \ref{fig1}(c), the
potential profile saturates at roughly $\pm 200\unit{nm}$, so we consider a
scattering region described by $H_{\text{bulk}}(V(x),t;k_{y})$
with length $L_{x}=400\unit{nm}$. The transport is solely supported by the
states at the global Fermi level, which is set to $E_{F}^{0}=E_{F}\left(
x=\pm 200\unit{nm}\right) $. We first investigate the top-gate voltage
dependence of the single-mode conductance $g$. In Fig.\ \ref{fig2}(a), we
directly compare the oscillating features of our computed $g$ with the
experimental data $G_{\text{YK}}$,\footnote{%
The experimental data compared in this work were extracted from the
electronic file of Ref.\ \onlinecite{Young2009}, instead of the original
data.} choosing the measured $G_{\text{YK}}(V_{\text{tg}},V_{\text{bg}}=40%
\unit{V})$ and $G_{\text{YK}}(V_{\text{tg}},V_{\text{bg}}=60\unit{V})$
curves as explicit examples. In both cases, the general features of the
measured oscillating conductance are well captured by our TBM calculation.
The Dirac point position of the locally-gated region corresponds to the
conductance dip. To the left of this minimum the transport is in the \emph{%
npn} regime exhibiting Fabry-P\'{e}rot-type oscillations due to interference
of backscattered waves between the \emph{np} and the \emph{pn} interfaces.
To the right of the dip, the transport enters the \emph{nn'n} regime, where
graphene becomes much more transparent than for \emph{npn}, resulting in the
suppression of the interference and the rise in the conductance. This
conductance asymmetry\cite{Huard2007,Stander2009,Cayssol2009,Low2009a} is
the first indirect feature of Klein tunneling, which results in the decay of
the transmission with the incident angle in the \emph{np} regime\cite%
{Cheianov2006} and hence a lower integrated conductance, although the
tunneling at normal incidence is perfect.

The single-mode spin-degenerate conductance $g$ from Eq.\ \eqref{g} has a
maximum of $2e^{2}/h$ and does not reflect the main effect of the back-gate
voltage that tunes the global Fermi level $E_{F}^{0}$: the modulation of the
number of modes $M$ participating in transport. For bulk graphene at low
energy, $M$ can be approximated by $2k_{F}/\Delta k_{y}$ with $\Delta
k_{y}=2\pi /L_{y}$, where $L_{y}$ is the width of the graphene flake. This
gives $M(E)=2L_{y}\left\vert E\right\vert /(\pi \hbar v_{F})$. While the
calculation considers the bulk transport across the locally gated region in
graphene, the contact resistance $R_{c}$ between the electrodes and graphene
is not included. To compare with the full map of the measured $G_{\text{YK}%
}(V_{\text{tg}},V_{\text{bg}})$, we temporarily adopt a simple model to
account for multiple modes and contact resistance: $G(E_{F}^{0})=%
\{[M(E_{F}^{0})g(E_{F}^{0})]^{-1}+R_{c}\}^{-1}.$ Assuming an effective width
$L_{y}=2\unit{%
\mu%
m}$ and a low contact resistance $R_{c}=0.2\unit{k%
\Omega%
}$, we display the calculated top- and back-gate dependencies of $%
G(E_{F}^{0})$ in Fig.\ \ref{fig2}(b), which qualitatively agrees with Ref.\ %
\onlinecite{Young2009}. Note that the quadrants of $G(V_{\text{tg}},V_{\text{%
bg}})$ are determined by the dependence of the potential profile on $V_{%
\text{tg}}$ and $V_{\text{bg}}$, and do not significantly change with the
temporarily introduced parameters $L_{y}$ and $R_{c}$, on which we place
less stress in the present work.

\subsection{Low-field magnetotransport}

Finally, we come to a closer analysis of the low-field magnetotransport. For
an \emph{incoherent} graphene $\emph{pnp}$ junction a perpendicular magnetic
field leads to the increase in the magnetoresistance due to the bending of
the electron trajectories.\cite{Cheianov2006} When the top-gate is narrow
enough, such as that in Ref.\ \onlinecite{Young2009}, with a width of about $%
20\unit{nm}$, a \emph{coherent }graphene \emph{pnp} junction can be formed.
Shytov \textit{et al.}\cite{Shytov2008}\ proposed a clever way to
experimentally test the existence of Klein tunneling, making use of the sign
change of the Klein backscattering phase at a weak magnetic field, which in
turn results in a half-period shift of the Fabry-P\'{e}rot oscillations.
Based on this semiclassical treatment the low-field magnetotransport
experiment in Ref.\ \onlinecite{Young2009} was regarded as providing
evidence of Klein tunneling. In the following we show that our
tuning-parameter-free TBM calculation confirms the semiclassical picture
and, again, agrees well with the measurement.

\begin{figure}[t]
\includegraphics[width=\columnwidth]{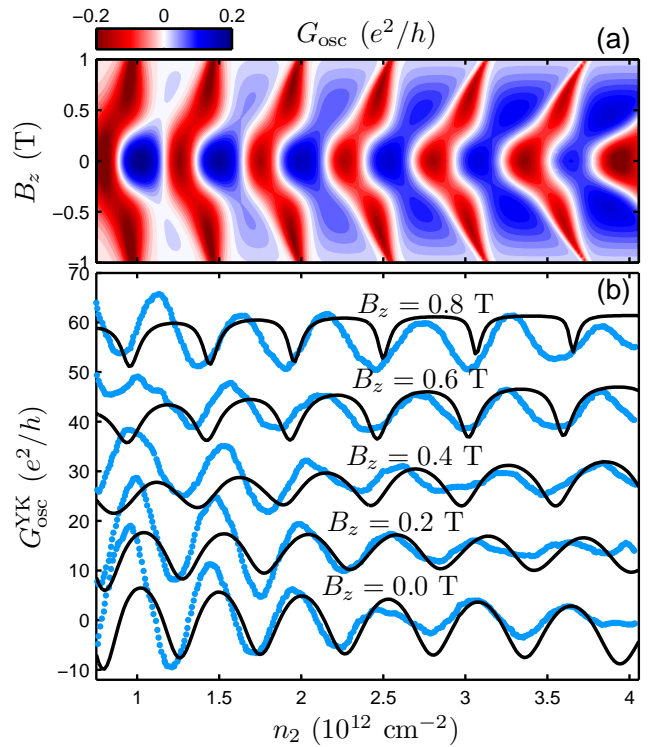}
\caption{{}(Color online) (a) Oscillating part of the computed conductance $%
G_{\text{osc}}(n_{2},B_{z})$ (see text for definition) as a function of the
carrier density of the locally gated region $n_{2}\equiv n(x=0)$ and the
external magnetic field $B_{z}$. (b) Comparison of computed $G_{\text{osc}}$
curves [solid (black) curves] at various magnetic field strengths with the
experimental data from Ref.\ \onlinecite{Young2009} [blue (gray) dots]
\lbrack dotted gray (blue) curves].}
\label{fig3}
\end{figure}

The orbital contribution of the external magnetic field $B_{z}$
perpendicular to the graphene plane is incorporated in the TBM calculation
through the Peierls substitution,\cite{Peierls1933} while the Zeeman term is
neglected since the Zeeman splitting is rather small compared to $E_{F}^{0}$.%
\cite{DasSarma2011} To maintain the transverse ($y$) translation invariance
throughout the whole system while also keeping the longitudinal ($x$)
translation invariance in the leads, we consider the Landau gauge of $%
\mathbf{A}=(0,xB_{z},0)$ only in the scattering region. Inside the left and
right leads, however, constant gauge field strengths $A_{y}^{L}=x_{L}B_{z}$
and $A_{y}^{R}=x_{R}B_{z}$ must be considered, respectively, where $x_{L}$
and $x_{R}$ are the position coordinates of the left-most and right-most
atomic site of the scattering region, in order to avoid a discontinuity of
the vector potential.

Since the expected phase shift stems from Klein backscattering between the
two interfaces inside the locally gated region, the potential tail does not
play a crucial role and we reduce the scattering region length to $L_{x}=150%
\unit{nm}$. Following the definition of the oscillating part of the
conductance given in Ref.\ \onlinecite{Young2009}, we process our data on
the single-mode conductance $g$ by first computing the odd part of the
conductance, $G_{\text{odd}}(n_{2},B_{z})=g(n_{2},B_{z})-g(-n_{2},B_{z})$,
and then subtracting its mean value to obtain $G_{\text{osc}%
}(n_{2},B_{z})=G_{\text{odd}}(n_{2},B_{z})-\overline{G_{\text{odd}%
}(n_{2},B_{z})}$. Here $n_{2}=n(x=0)$ [see Eq.\ \eqref{n(x)}] is the carrier
density of the locally-gated region. The obtained oscillation fringes of $G_{%
\text{osc}}(n_{2},B_{z})$ are shown in Fig.\ \ref{fig3}(a), which is, again,
qualitatively consistent with Ref.\ \onlinecite{Young2009}. The sudden phase
shift, which indicates the presence of perfect transmission and corresponds
to the half-period shift predicted by Shytov \textit{et al.},\cite%
{Shytov2008} occurs at magnetic field strengths between $0.2\unit{T}$ and $%
0.4\unit{T}$ and is in excellent agreement with Ref.\ \onlinecite{Young2009}%
. In Fig.\ \ref{fig3}(b), the computed $G_{\text{osc}}$ is compared with the
experimental data $G_{\text{osc}}^{\text{YK}}(n_{2},B_{z})$\cite{Note1} at
various magnetic field strengths (both with offset for clarity).

\section{Summary\label{sec summary}}

In conclusion, we have demonstrated the applicability of TBM-based quantum
transport simulations for transport in bulk graphene heterojunctions.
Applying the Bloch theorem along the transverse dimension, the computational
effort for TBM transport through a bulk scattering region is significantly
reduced. Together with the realistic potential profile extracted from the
carrier density profile of a graphene \emph{pnp} junction, this method
provides a confirmation of the experiment in Ref.\ \onlinecite{Young2009}
and its semiclassical theoretical interpretation, at a low computational
cost without using free tuning parameters. The quantum transport approach
presented here for studying bulk properties is suitable not only for
graphene but also for other materials where the TBM works well.

\begin{acknowledgments}
We appreciate valuable discussions with A.\ Cresti and V.\ Krueckl.
Financial support from the Alexander von Humboldt Foundation (M.-H.L.) and
Deutsche Forschungsgemeinschaft within GRK1570 (K.R.) is gratefully
acknowledged.
\end{acknowledgments}

\bibliographystyle{apsrev4-1}
\bibliography{mhl2}

\end{document}